\documentclass[useAMS,usenatbib]{mn2e}
\usepackage{aecompl}
\usepackage{amsmath}
\usepackage{bethmacros}
\usepackage{booktabs}
\usepackage{color}
\usepackage[T1]{fontenc}
\usepackage{graphicx}
\usepackage{hyperref}
\usepackage{xspace}

\newcommand\msun{\ensuremath{{\rm M}_\odot}\xspace}
\def \eg {e.\,g.\xspace}
\def \ie {i.\,e.\xspace}
\def\fb{\ensuremath{f_\textrm{b}}\xspace}
\newcommand\rtwo{\ensuremath{R_{200}}\xspace}
\newcommand\mtwo{\ensuremath{M_{200}}\xspace}

\newcommand\mgas{\ensuremath{M_{\rm gas}}\xspace}
\newcommand\mstar{\ensuremath{M_\star}\xspace}
\newcommand\sig[1]{\ensuremath{\Sigma_{\rm #1}}\xspace}
\newcommand\radius[1]{\ensuremath{R_{\rm #1}}\xspace}

\newcommand\changa{{\sc ChaNGa}\xspace}
\newcommand\gasoline{{\sc gasoline}\xspace}

\title[Halo spin bends galaxy disc profiles]{How to bend galaxy disc profiles: the role of halo spin}
\author[Herpich et al.]{J. Herpich$^1$\thanks{Email: herpich@mpia.de}\thanks{Member of the International Max Planck Research School for
Astronomy and Cosmic Physics at the University of Heidelberg,
IMPRS-HD, Germany.},
G.\,S. Stinson$^{1}$, A.\,A. Dutton$^{1}$, H.-W. Rix$^{1}$, M. Martig$^{1}$,
\newauthor{R. Ro\v skar$^{2}$, A.\,V. Macci\`o$^{1}$,
T.\,R. Quinn$^{3}$, J. Wadsley$^{4}$}
\vspace*{6pt}\\
$^{1}$Max-Planck-Institut f\"ur Astronomie, K\"onigstuhl 17, D-69117 Heidelberg, Germany\\
$^{2}$Research Informatics, Scientific IT Services, ETH Zurich, Weinbergstrasse 11, 8092 Zurich, Switzerland\\
$^{3}$Astronomy Department, University of Washington, Box 351580, Seattle, WA 98195-1580, USA\\
$^{4}$Department of Physics and Astronomy, McMaster University, Hamilton, ON L8S 4M1, Canada\\
}

\begin{document}
\maketitle

\begin{abstract}
The radial density profiles of stellar galaxy discs can be well approximated as an exponential.
Compared to this canonical form, however,  the profiles in the majority of disc galaxies
show downward or upward breaks at large radii. Currently, there is no coherent explanation
in a galaxy
formation context of the radial profile per se, along with the two types of profile breaks.
Using a set of controlled hydrodynamic simulations of disc galaxy formation, we
find a correlation between the host halo's initial angular momentum and the resulting
radial profile of the stellar disc: galaxies that live in haloes with a low spin parameter
$\lambda\la0.03$ show an up-bending break in their
disc density profiles, while galaxies in haloes of higher angular momentum
show a down-bending break. We find that the case of pure exponential profiles
($\lambda\approx0.035$)
coincides with the peak of the spin parameter distribution from cosmological
simulations. Our simulations not only imply an explanation of
the observed behaviours, but also suggest that the physical origin of
this effect is related to the amount of radial redistribution of stellar mass,
which is anti-correlated with $\lambda$.
\end{abstract}


\section{Introduction}
\label{sec:intro}

In his landmark work, \citet{Freeman1970} found that most spiral galaxies share a uniform
stellar surface brightness profile which is well fitted by an exponential,
$\mu_\star\left(R\right)\propto\exp\left(-R/\radius d\right)$.
Subsequent deeper imaging of a wider variety of disc galaxies has shown some variation in the
functional form of the radial profile.
For example \citet{Pohlen2002} observed galaxies %
that exhibit two distinct exponential profiles.
In addition to an inner profile just as observed by \citet{Freeman1970}, they found a steeper
outer exponential profile such that the overall profile appeared to have a ``break''.
In barred S0-Sb galaxies
\citet{Erwin2005} found profiles with an outer exponential slope which is shallower than the
inner part.
\citet{Pohlen2006} compiled a sample of galaxies that included all three types of profiles.
They called them \emph{pure, down-bending} and \emph{up-bending} exponentials.
They found that only $\approx10\%$ of their sample of
late-type galaxies have pure exponential profiles that extend all the way out to the
observational surface brightness limit.
The measured abundances of down- and up-bending profiles are $\approx60\%$ and $\approx30\%$,
respectively.

Using numerical simulations, \citet{Debattista2006} explain down-bending disc breaks
with stellar angular momentum redistribution induced by the formation of a bar.
A study by \citet{Roskar2008} that also employed numerical simulations predicted
`U-shaped' stellar age profiles for down-bending disc breaks and that the positions
of the minima of the age profiles coincide with the break radius.
This prediction was confirmed by observations \citep{Yoachim2010}.

There are many analytical models in the literature that study the physical origin of
the exponential radial profile of stellar galaxy discs
\citep[\eg][]{Fall1980, Lin1987, Dalcanton1997, Mo1998, vandenBosch2001}.
A common assumption of all these models is that the distribution of specific angular
momentum of baryons is conserved during the evolution of galaxies.
\citet{Dutton2009} found that low angular momentum material needs to be removed in
order to prevent centres of galaxies from becoming too dense.
He suggested that stellar feedback is a viable mechanism to redistribute angular momentum.
Cosmological simulations of galaxy formation confirm this mechanism and find that
this ejected low angular momentum gas can be reaccreted with high angular momentum
via the \emph{galactic fountain} effect \citep[\eg][]{Brook2011, Brook2012, Marinacci2011,
Uebler2014}.
{However, currently there is no coherent explanation for the existence of pure, down- and
up-bending radial profiles.}

In this Letter we present the first attempt at identifying physical quantities that
determine the profile of stellar discs at large radii.
We use a suite of smoothed particle hydrodynamic (SPH) simulations of
disc galaxies in a{{}n isolated} set-up.
We show that the type of disc profiles depends on the initial angular momentum of the
galactic baryons.
In Section \ref{sec:simulations}, we describe the simulation set-up followed by a presentation
of the results in Section \ref{sec:results} and our conclusions in Section \ref{sec:conclusions}.

\section{Simulations}
\label{sec:simulations}

In this Letter we present the results of a suite of simulations of disc galaxy formation
from idealized and { isolated}, yet cosmologically motivated, initial conditions.
We use a modified version of the publicly available treeSPH code \changa
\citep{Jetley2008, Jetley2010, Menon2014}
\footnote{\url{http://librarian.phys.washington.edu/astro/index.php/Research:ChaNGa}}.
The simulations are evolved for 8 Gyr.
This corresponds to $z\sim1.5$, when the last major merger era was coming
to an end \citep{Zentner2003}.
Examinations of cosmological simulations show that they behave similarly to isolated
spheres after their last major merger \citep[][Obreja et al., in preparation]{Zentner2003}.
In section \ref{sec:ics} we present a detailed description of our cosmologically motivated
initial set-up while
we will only briefly outline the implementation of hydrodynamics in section
\ref{sec:hydro}.

Our initial set-up is motivated and indeed very similar to that in \citet{Roskar2008}.
We deliberately chose a simplified and controlled set-up in order to be able to link
observed properties to their physical origin more easily.
This level of control comes at the expense of neglecting asymmetric influences %
from the cosmological context, such as merging of galaxies. %
{ However, stellar feedback helps to break
the symmetry in the initial conditions.}

\subsection{Initial conditions}
\label{sec:ics}

We set up isolated haloes with the following properties: $\mtwo = 10^{12}\,\msun,
\rtwo = 206\,\textrm{kpc}, f_\textrm{b} = 0.1, c=10$.
\mtwo and \rtwo are the virial mass and radius. %
$f_\textrm{b}$ is the fraction of baryons in the initial set-up and $c$ is the halo
concentration.
{\fb is lower than the cosmological baryon fraction because our set-up does not account
for high redshift outflows that are ejected from the halo.
Dark matter (DM) particle masses are $1.1\times10^6\,\msun$ and initial gas particle masses
are $1.2\times10^5\,\msun$.
The gravitational softening is $\epsilon=227\,\textrm{pc}$. %
The SPH smoothing length is variable and set such that the kernel covers 50 particles.}

The initial conditions were set up in four steps.
First we created an equilibrium NFW DM halo \citep{NFW} following the recipe from
\citet{Kazantzidis2004} including an exponential cutoff outside \rtwo.
In the next step the mass of each DM particle was reduced by a factor of \fb,
the baryon fraction,
and a gas particle was added at the same position, accounting for the mass difference
between the old and the new DM particle.

This gas sphere is then rotated by some random angle in order to prevent gas and DM
particles from sharing identical positions.

To set the gas velocities, we establish a cylindrical coordinate system
$(v_\textrm{R}, v_\textrm{c}, v_\textrm{z})$ such that the gas orbits about the z-axis.
The velocities are set to obey the angular momentum profile for DM haloes
as found by \citet{Bullock2001} in cosmological $N$body simulations%
\footnote{These calculations were done assuming that the DM and the gas share the same
angular momentum profile.}:
\begin{equation}
\frac{M\left(<j\right)}\mtwo = \mu\frac{j/j_\textrm{max}}{j/j_\textrm{max}+\mu-1},
\label{eq:angmom_profile}
\end{equation}
where $M\left(<j\right)$ is the mass of all material that has less angular momentum than $j$,
$\mu$ is the shape parameter and $j_\textrm{max}$ is the maximum specific angular momentum
in the halo.
$j_\textrm{max}$ depends on the value of $\mu$ and is proportional to the spin parameter
$\lambda$.
{ While $\lambda$ simply scales the gas particles' angular momentum, $\mu$ sets the
actual mass distribution of $j/j_\textrm{max}$.}
We use the definition of the spin parameter from \citet{Bullock2001}:
\begin{equation}
\lambda = \left. \frac J {\sqrt{2}MVR}\right|_{R=\rtwo}
\end{equation} 
Here $J$ and $M$ are the halo angular momentum and mass inside a sphere of radius $R$ and
$V$ is the halo circular velocity at that radius.
Radial and vertical velocities were set to $v_\textrm{R}=v_\textrm{z}=0$.
Tangential velocities are a function of the axisymmetric radius only
($v_\textrm{t}\left(R,\phi,z\right)=v_\textrm{t}\left(R\right)$).
{ However, the DM halo does not rotate in our simulations.}
Finally, the gas temperatures were calculated such that the gas obeys hydrostatic
equilibrium.

Here we explore the effects of varying %
$\lambda$ at a fixed $\mu=1.3$ on the radial profile of the resulting stellar disc.
We explore the range $0.02\le\lambda\le0.1$\footnote{We did not explore lower values of
$\lambda$ because the computational effort increases significantly as $\lambda$ decreases
due to denser gaseous discs and an increased amount of star formation.}.
The simulation parameters are summarized in Table \ref{tab:simulation_pars}.
\begin{table}
    \caption[Simulation parameters]{
        Overview of all simulations and their properties.
        $\lambda$ is the initial spin parameter, \mgas and \mstar are the amount of
        gas and stars in the disc region ($R<30\,{\rm kpc}, \left|z\right|<3\,{\rm kpc}$)
        at 8 Gyr.
    }
    \label{tab:simulation_pars}
    \begin{tabular}{ccc}
    \toprule
    $\lambda$ & \mgas & \mstar \\
        & ($10^{10}$ \msun) & ($10^{10}$ \msun) \\
    \midrule
    0.02 & 0.34 & 3.14 \\
    0.03 & 0.56 & 2.91 \\
    0.035 & 0.67 & 2.64 \\
    0.04 & 0.95 & 2.53 \\
    0.045 & 1.13 & 2.42 \\
    0.05 & 1.24 & 2.36 \\
    0.055 & 1.31 & 2.30 \\
    0.06 & 1.34 & 2.25 \\
    0.1 & 1.44 & 1.96 \\
    \bottomrule
    \end{tabular}
\end{table}

\subsection{Baryonic physics}
\label{sec:hydro}
The \changa code is derived from the treeSPH code \gasoline.
It uses a modified version of SPH which employs a pressure averaged force calculation
\citep{Ritchie2001, Hopkins2013, Keller2014}.
\changa includes stochastic star formation \citep[$c_\star=0.1$]{Stinson2006}
based on a Kennicut-Schmidt law, radiative metal line cooling,
metal diffusion and pre supernova stellar wind feedback
\citep[early stellar feedback;][]{Stinson2013}.
The feedback follows \citet{DallaVecchia2012} in which
the energy output from supernova explosions of a stellar population is released at one time
altogether. %
The energy released per supernova is $E_\textrm{SN}=1.5\times10^{51}\,\textrm{erg}$.
Further details on the implemented physics in \changa will be presented in upcoming papers
\citep[Stinson et al., in preparation]{Keller2014}.
{ First tests of this implementation produced realistic disc galaxies in cosmological
simulations.}

\section{Results}
\label{sec:results}

\subsection{Disc profiles}
\label{sec:disc_profiles}
\begin{figure}
\includegraphics[width=\columnwidth]{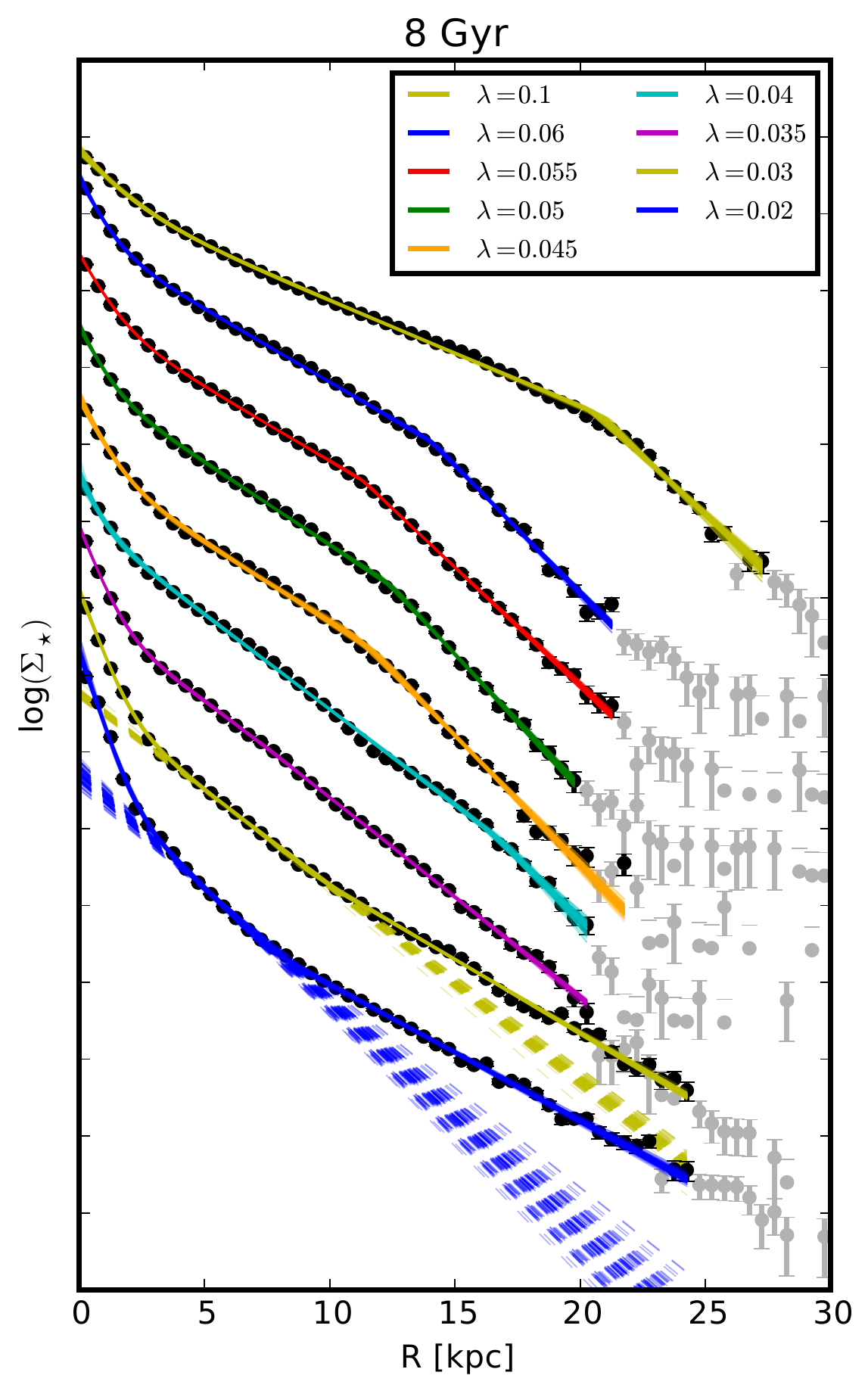}
\caption{
    The radial stellar surface density profiles.
    Presented is the stellar surface density as a function of axisymmetric radius
    of the individual haloes at $t=8$ Gyr.
    { The errors are estimated as Poisson noise.}
    The individual profiles are offset by 1 dex (\ie one tick mark) each for
    clarity as we are not interested in the normalization.
    The grey data points correspond to radial bins with nine or less star particles
    which were ignored in the fitting procedure.
    { The coloured lines show the model for 100 sets of parameters which were randomly
    chosen from the PDF obtained from the MCMC algorithm.}
    For the up-bending disc profiles, the dashed lines are an extrapolation of
    the inner exponential part ($\Sigma_0\exp\left(-R/\radius i\right)$) to make the
    break in the profile more easily visible.
    The figure qualitatively shows that there is a trend from up-bending disc profiles for
    galaxies in low spin haloes to down-bending profiles in high spin haloes.
    \label{fig:profiles_and_fits}
}
\end{figure}
After evolving the simulations for 8 Gyr, we extracted the stellar surface density
in axisymmetric radial bins $\sig\star\left(R\right)$.
These stellar surface density profiles include all stars up to 3 kpc above and below
the plane and inside a cylinder with a radius of 30 kpc.
We fit these profiles with a superposition of a `broken exponential disc' $\sig d$
surrounding an exponential bulge which is a very good parametrization of the data
(see Fig. \ref{fig:profiles_and_fits}):
\begin{equation}
\sig\star\left(R\right) = \sig k\exp\left(-\frac R{\radius k}\right)+ \sig d\left(R\right)
\label{eq:expbulge_broken_exponential}
\end{equation}
where
\begin{equation}
\sig d\left(R\right) = \Sigma_0\times
    \begin{cases}
        \exp\left(-\frac R{\radius i}\right) & \textrm{if}\ R<\radius b \\
        \exp\left(-\frac {\radius b}{\radius i}\right)
            \exp\left(-\frac{R-\radius b}{\radius o}\right) & \textrm{else}
    \end{cases}
\label{eq:broken_exponential}
\end{equation}
Here \sig k and \sig0 are normalization factors for the bulge and disc component, respectively.
\radius i and \radius o represent the inner and outer disc scale-length and \radius b
the radius of the break.
{ The probability distribution function (PDF) of the fit parameters for the given surface
density profiles was obtained using the Monte Carlo Markov chain (MCMC) algorithm
{\sc emcee} \citep{emcee}}\footnote{We did not use a standard
$\chi^2$ minimization procedure because it gave unstable results.}.
Fig. \ref{fig:profiles_and_fits} shows the stellar surface density profiles (black points)
for all simulations at 8 Gyr overplotted by {100 models sampled from the
obtained PDF (coloured lines).}
The grey data points show data for radial bins with nine or fewer star particles.
These points were omitted in the fitting procedure.

\subsection{Disc breaks}
\label{sec:disc_breaks}
\begin{figure}
\includegraphics[width=\columnwidth]{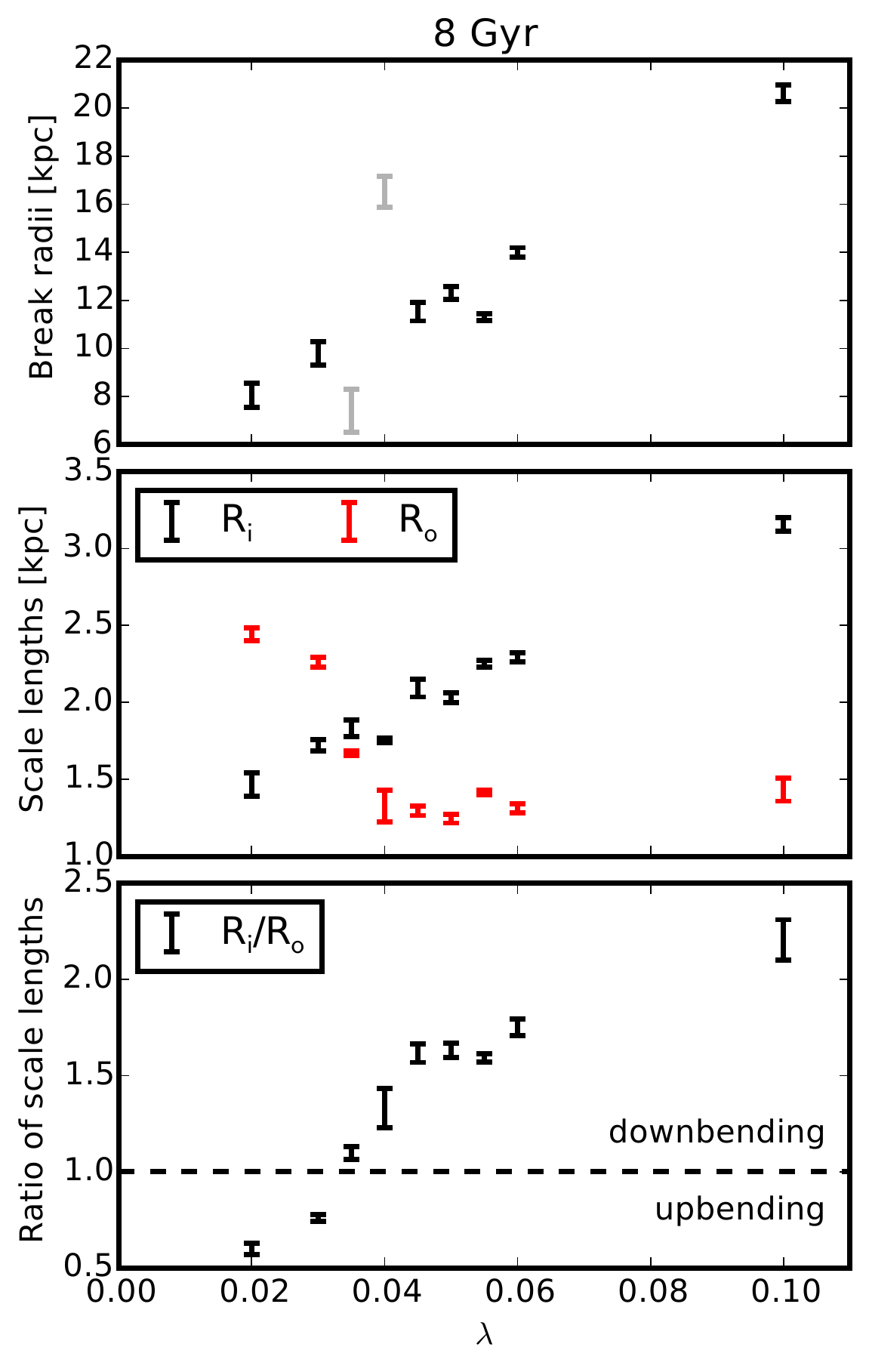}
\caption{
    Systematic variation of the parameters for the broken exponential disc profile
    (equation \eqref{eq:broken_exponential}) as a function of $\lambda$.
    The top panel shows the break radii.
    The estimates for the break radius for $\lambda=0.035, 0.04$ are plotted in grey
    since they are somewhat ill-defined in these cases (see section \ref{sec:pure_exponential}).
    The middle panel shows the estimated inner (black) and outer (red) scale radii.
    The bottom panel shows the ratio of the inner and outer scale-lengths.
    The horizontal dashed line indicates unity, that is the separation between the
    up- (below the line) and down-bending regime (above).
    The error bars indicate the range between the 16-th and the 84-th percentile.
    The figure shows a clear trend from up- to down-bending profiles as $\lambda$ increases
    and a transition at $\lambda\approx0.035$.
    \label{fig:fitted_radii}
}
\end{figure}

Fig. \ref{fig:profiles_and_fits} shows that the type of the disc profiles
changes with $\lambda$.
In the lowest spin simulations ($\lambda\le0.03$) the profile is up-bending,
\ie $\radius o>\radius i$.
The models with high spin parameters ($\lambda\ge0.045$) clearly show a down-bending
break.
Fig. \ref{fig:fitted_radii} quantifies these trends in terms of the fit parameters.
The middle panel shows the fitted inner to outer scale-lengths
as a function of $\lambda$.
The inner scale-length \radius i increases linearly with $\lambda$.
The outer scale-length \radius o has a high value for low $\lambda$ and decreases with $\lambda$
until it approaches a constant lower value for $\lambda\ga0.04$.
The bottom panel of Fig. \ref{fig:fitted_radii} shows the ratios of the inner
to outer scale-lengths $\radius i/\radius o$ as a function of $\lambda$.
There is a transition from up-bending ($\radius i/\radius o<1$) to down-bending
($\radius i/\radius o>1$) disc breaks at $\lambda\approx0.035$.

The top panel in Fig. \ref{fig:fitted_radii} shows the estimated break radii \radius b as a
function of $\lambda$.
Except for the $\lambda=0.035, 0.04$ cases, \radius b shows a linear dependence on $\lambda$.

\subsection{The pure exponential case}
\label{sec:pure_exponential}
In the region between the up- and down-bending regime ($\lambda=0.035-0.04$), the disc profiles
are purely exponential over a range that exceeds the typical position of disc breaks
(8-10 kpc) for other spin parameters.
In such cases, the definition of \radius b and the distinction between \radius i and
\radius o are ill-defined.
In the $\lambda=0.04$ case, the best fit for \radius b is at the end of the disc.
Therefore, the fitted break radius exceeds the otherwise linear relation with $\lambda$ as
can be seen in the middle panel of Fig. \ref{fig:fitted_radii}.
In the $\lambda=0.035$ case, %
the best fit for \radius b is at a minor wiggle in the profile.
The result is that \radius b does not fit the relation with $\lambda$ either but this time
it falls short.
{ This is not surprising given that the model parameters are degenerate for
pure exponential profiles where
\radius b and the distinction between \radius i and \radius o have no
physical meaning.
This phenomenon is reflected by the larger error bars.
}

\subsection{Radial mass redistribution}
\label{sec:redistribution}
\begin{figure}
\includegraphics[width=\columnwidth]{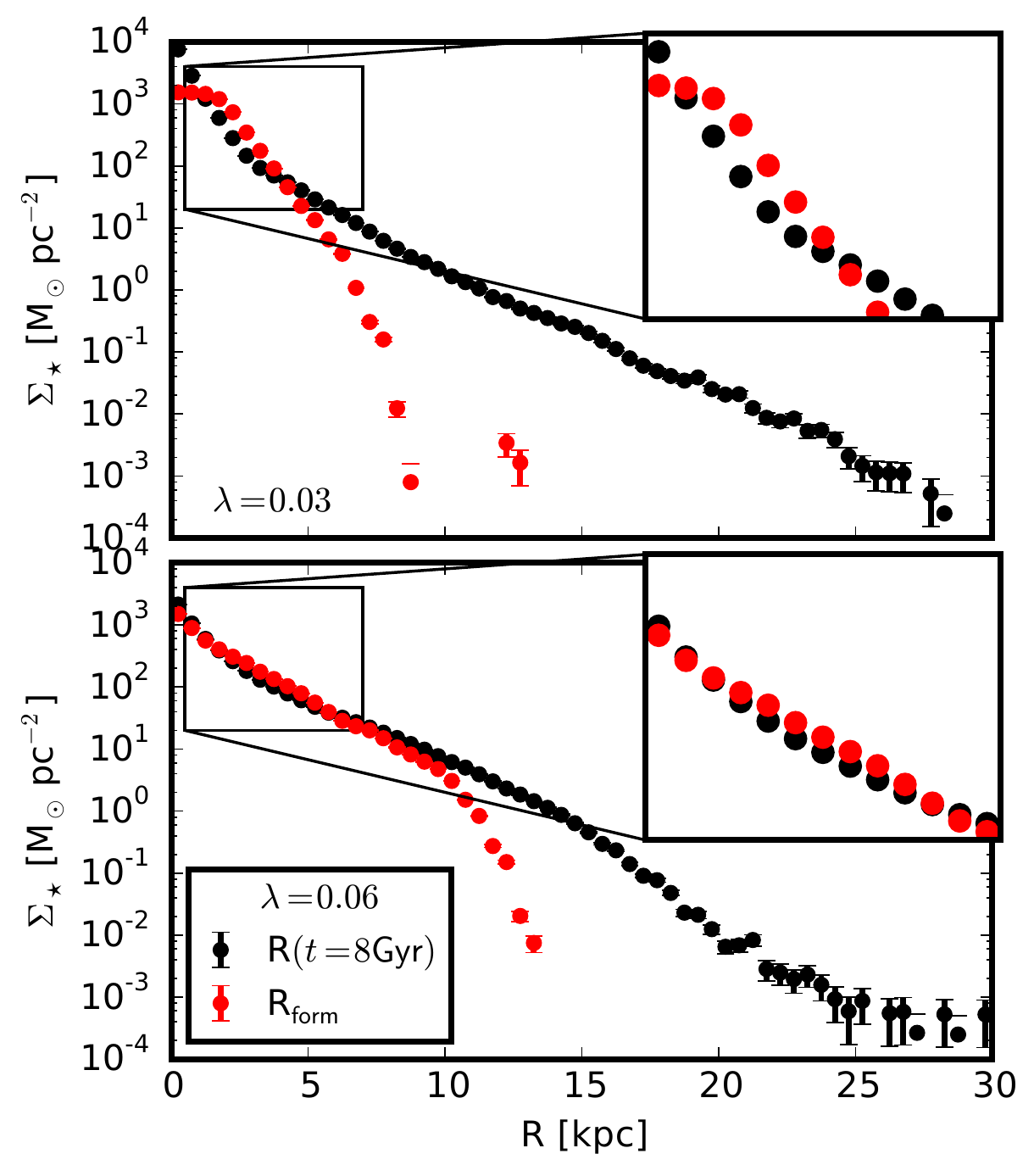}
\caption{
    Comparison of stellar surface density profiles relative to the current position
    of star particles (black) and their position at the time of formation (red).
    The top and bottom panels show the results for $\lambda=0.03$ and $0.06$,
    respectively.
    The figure clearly shows that the amount of radial redistribution of stellar mass
    is much more pronounced for the case with small $\lambda$.
    \label{fig:current_birth_radii}
}
\end{figure}

In this section we briefly lay out a possible mechanism that may be the cause for
systematic changes of the disc profile with $\lambda$.
Fig. \ref{fig:current_birth_radii} compares the final stellar surface density profile
to the density profile of stars at birth, irrespective of their actual time of birth.
Hereafter, we will refer to these as the \emph{final} and \emph{formation} profiles
respectively.

The two panels in Fig. \ref{fig:current_birth_radii}
show profiles for two different simulations, one with a spin parameter
below the transition region between up- and down-bending discs ($\lambda=0.03$, top panel)
and one above ($\lambda=0.06$, bottom panel).
In the $\lambda=0.03$ case we see that stars are formed with a very
concentrated profile that has a core in the centre.
This core might be  artificial due to
centring issues.
What is evident without doubt is that practically all stars formed in the inner $\approx 8$ kpc
of the disc, while the final profile extends all the way out to $\approx30$ kpc.

The figure shows that in the range $2\,{\rm kpc} \la R \la 5\,{\rm kpc}$
the formation profile significantly exceeds the final profile.
Inside and outside that range the opposite is the case.
This is because the integrated masses of the current and formation profiles are identical.
It follows that a substantial amount of stellar redistribution has occurred
from $2\,{\rm kpc} \la R \la 5\,{\rm kpc}$ outwards as well as inwards.

For $\lambda=0.06$, we observe the same qualitative effect but there are two striking
differences with important implications.
First, star formation extends to much larger radii ($\approx 13$ kpc).
Secondly, despite the same qualitative trend the differences between the formation and final
profiles are much smaller.
The final profile closely follows the formation profile out to $\approx 10$ kpc and
there is no central core in the formation profile.
{ At large radii there is a steeply declining tail of stars that have reached large radii,
leading to the down-bending profile.}
Therefore the amount of stellar redistribution that took place in this `high-spin' case
is much less than that in the `low-spin' case.
For clarity, we only presented these two cases.
The intermediate cases show a monotonic trend between those shown in Fig.
\ref{fig:current_birth_radii}.

\section{Conclusions}
\label{sec:conclusions}
Using numerical models of disc galaxy formation,
we found a correlation between the initial spin of the host halo and the shape of the radial
profile of the stellar disc.

We find that galaxies with an initial spin parameter $\lambda\la0.035$
form an up-bending disc profile while larger values of $\lambda$ yield down-bending discs.
Pure exponential discs occur only right at $\lambda\approx0.035$
{ which} coincides approximately with the median of the
{ $\lambda$} distribution
in cosmological simulations \citep[$\lambda=0.031$;][]{Maccio2008}.
Thus, our model explains why only some late-type galaxies
exhibit pure exponential disc profiles, while the majority of them show breaks in the outer
disc profiles.
As the transition between up- and down-bending disc profiles ($\lambda=0.035$) in our model
coincides with the median of the {$\lambda$} distribution, we expect roughly equal
abundances
of up- and down-bending profiles which are comparable but not equal to observational results
\citep[30 \% and 60 \%;][]{Pohlen2006}.
{ A possible cause for this discrepancy between our model and observations is that
gas in cosmological simulations generally has larger angular momentum than the DM
\citep{Stewart2013}.}

Nevertheless the model qualitatively reproduces all types of observed disc profiles.
In a follow-up paper we will further explore this effect in cosmological zoom simulations.

\section*{Acknowledgements}
The authors thank Frank van den Bosch for his very helpful comments on setting up
the simulations.
The analysis was performed using the {\sc pynbody} package
\citep{pynbody}.
The simulations were performed on the \textsc{THEO} cluster of  the
Max-Planck-Institut f\"ur Astronomie and the Hydra supercomputer of the
Max-Planck-Gesellschaft at the Rechenzentrum in Garching.
The authors greatly appreciate the contributions of these computing allocations.
JH, GSS and HWR acknowledge funding from the European Research Council under the
European Union's Seventh Framework Programme (FP 7) ERC Advanced Grant Agreement no. [321035].
GSS, AAD and AVM acknowledge support through the  Sonderforschungsbereich SFB 881
`The Milky Way System' (subproject A1) of the German Research Foundation (DFG).
MM acknowledges support from the Alexander von Humboldt Foundation.

\end{document}